\documentclass[
aps,
10pt,
prl, 
twocolumn,
letterpaper, 
superscriptaddress]
{revtex4-2}
\usepackage[dvipsnames]{xcolor}
\usepackage{amsmath, amssymb}
\usepackage{hyperref, cleveref}
\usepackage{graphicx}

\newcommand{\mb}[1]{\mathbf{#1}}

\begin{document}
\title{Enhanced enantiomer discrimination with chiral surface plasmons}

\author{Sang Hyun Park}
\affiliation{Department of Electrical \& Computer Engineering, University of Minnesota, Minneapolis, MN, USA}
\affiliation{Department of Physics, The University of Texas at Austin, Austin, TX, USA}
\author{Phaedon Avouris}
\affiliation{Department of Electrical \& Computer Engineering, University of Minnesota, Minneapolis, MN, USA}
\author{Jennifer A. Dionne}
\affiliation{Department of Materials Science and Engineering, Stanford University School of Engineering, Stanford, CA, USA}
\affiliation{Department of Radiology, Stanford University School of Medicine, Stanford, CA, USA}
\author{Joshua D. Caldwell}
\affiliation{Interdisciplinary Materials Science Graduate Program, Vanderbilt University, Nashville, TN, USA}
\affiliation{Vanderbilt University, Department of Mechanical Engineering, Nashville, TN, USA}
\author{Tony Low}
\email{tlow@umn.edu}
\affiliation{Department of Electrical \& Computer Engineering, University of Minnesota, Minneapolis, MN, USA}

\begin{abstract}
Strong light-matter coupling in chiral cavities has been proposed as an effective way to selectively interact with an enantiomer that shares the same handedness as the cavity’s chiral mode. We show that surface plasmons supported by a two-dimensional interface with both electric and chiral conductivities discriminate enantiomers more efficiently than chiral optical cavities. A quantum-electrodynamic treatment is developed to incorporate the molecule’s electric and magnetic dipole moments. We show that the discrimination factor for a chiral plasmon can exceed that of the best chiral-mirror cavity by almost an order of magnitude due to stronger field confinement. In addition, surface plasmons couple to a dipole’s projection onto an entire plane, whereas cavity (or free-space) modes couple only to a single polarization axis. This geometric difference produces a $\sqrt{2}$ orientation-averaged boost in chiral discrimination for chiral surface platforms. A handedness-preserving reflector further amplifies the enhancement, opening a practical route towards chiral sensing using twisted-layer platforms.
\end{abstract}
\maketitle


Chiral molecules (CMs) exist as enantiomers, which are nonsuperimposable mirror images of one another. It is well known that important pharmaceutical properties of a CM, such as its toxicity or potency, are tied to its handedness\cite{Hutt1996}. Hence, accurate and efficient detection of the chirality of a molecule is critical. The conventional method of detecting a molecule's chirality is through circular dichroism measurements\cite{Rodger1997}. However, due to a mismatch in molecule size and the wavelength of free-space light, the interaction between individual molecules and circularly polarized light is extremely weak. High molecular concentrations or long signal acquisition times are thus required to generate a detectable signal\cite{Kelly2005}. 

Utilizing a cavity that confines light into small mode volumes ($V$) is a well-known strategy to increase the light-matter interaction strength\cite{Walther2006}. When the coupling strength is strong enough to exceed dissipation rates, the light-matter coupling enters the strong coupling regime in which coherent energy transfer between light and matter results in hybrid light-matter states called polaritons. Strongly coupled polaritonic states in cavities have been used to modify chemical reactions\cite{Thomas2019}, demonstrate Bose-Einstein condensation\cite{Kasprzak2006}, and enhance non-linear effects\cite{Kena-Cohen2010}. Recently, single-handed chiral cavities formed using chiral mirrors have been proposed as a platform for exploring strong coupling with chiral molecules, i.e., chiral polaritonics \cite{Schafer2023, Riso2023, Riso2024}. Most notably, it has been shown that the avoided crossing between polaritonic branches of the cavity-molecule coupled system has a size that is dependent on the chirality of both the cavity and the molecule.

The mode volume of mirror-based cavities is, however, ultimately restricted by the diffraction limit $V\approx(\lambda/n)^3$, where $\lambda$ is the wavelength of light in vacuum and $n$ is the refractive index. Alternatively, plasmonic cavities are capable of confining light into length scales below the diffraction limit, where the plasmon length scale ($L$) can be 2 to 3 orders of magnitude smaller than the wavelength of propagating light ($\lambda$) at the same frequency\cite{Iranzo2018}. The enhanced coupling strength originating from small mode volumes places less stringent restrictions on the losses of the cavity, thus allowing strong coupling in plasmonic cavities with larger losses when compared to dielectric cavities\cite{Chikkaraddy2016, Baranov2018}. The same principles equally apply to the case of surface plasmons for which strong coupling has been extensively studied \cite{Gonzalez-Tudela2013,Tormo2015}. While there have been studies examining the interaction between CMs and plasmons in chiral metasurfaces\cite{Zhao2017}, such platforms typically provide an electromagnetic field with non-uniform chirality\cite{Schaferling2012} that can significantly weaken chiral optical signals from the molecule.

In this Letter, we develop a theoretical framework describing how chiral surface plasmons with spatially uniform optical chirality supported by twisted two-dimensional materials interact with chiral molecules. Twisted two-dimensional materials, most notably twisted bilayer graphene, have been studied as a platform for chiral surface plasmons in a number of recent works\cite{Lin2020, Stauber2020}. Here, we introduce a quantization scheme for the chiral surface plasmon and derive an analytic expression for the coupling strength with CMs. The chiral discrimination factor, defined as the difference in coupling strength of a chiral mode to CMs of opposite handedness, is calculated for a chiral surface plasmon mode coupled to a single CM with arbitrary orientation and an ensemble of $N$ randomly oriented CMs. We show that the discrimination factor can exceed that of chiral-mirror-cavity by almost an order of magnitude. Lastly, we establish design rules for reflector-enhanced platforms and demonstrate that only a chiral mirror can simultaneously tighten field confinement and preserve optical chirality, thereby yielding an additional few-fold boost in the discrimination factor.


\begin{figure}
      \centering
      \includegraphics{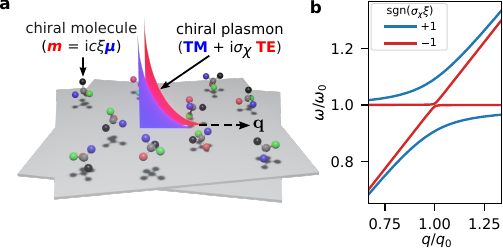}	
      \caption{General setup for coupling between chiral surface plasmons and chiral molecules. \textbf{a} Schematic representation of chiral molecules interacting with a chiral surface plasmon supported by a twisted 2D material. The chiral molecule is represented in the dipole approximation where the electric dipole $\boldsymbol{\mu}$ and magnetic dipole $\mathbf{m}$ are parallel with proportionality $ic\xi$. The chiral plasmon has a hybrid TM-TE field originating from the chiral conductivity $\sigma_\chi$ of the twisted atomic bilayer. \textbf{b} A schematic representation of the chirality-dependent energy level splitting for the coupled plasmon-CM system where $q_0$ is the plasmon wavevector at frequency $\omega_0$.}
      \label{fig:1}
\end{figure}

The general setup is shown in \cref{fig:1}a. A chiral surface plasmon mode with both transverse magnetic(TM) and transverse electric(TE) components propagates with an in-plane momentum $\mathbf{q}$. The handedness of the plasmon originates from the quantum interlayer interaction of the twisted bilayer material\cite{Stauber2018a}, which can be modeled as a bianisotropic surface conductivity with electric ($\sigma_e$) and chiral ($\sigma_\chi$) components\cite{Pfeiffer2014, Lin2020}. The chiral molecule is represented in the dipole approximation with both an electric $(\boldsymbol{\mu})$ and magnetic $(\mathbf{m})$ dipole moment. We assume that $\mb{m}=ic\xi\boldsymbol{\mu}$ since only the component of $\mb{m}$ parallel to $\boldsymbol{\mu}$ contributes to the molecule's chirality. The plasmon interacts with CMs of chirality $\xi$ near the surface of the twisted bilayer material. The coupling strength enters the strong coupling regime beyond a critical density of CMs\cite{Gonzalez-Tudela2013, Tormo2015}. Using a quantum electrodynamic approach\cite{Ferreira2020}, we analyze how the coupling strength between the plasmon and CM depends on their relative handedness. The coupling strength is manifested as the size of an avoided crossing (energy gap) in the spectrum of the collective system (see \cref{fig:1}b). Although the theory we develop is general, we will be focusing on the mid-infrared frequency range and thus fix the molecular vibrational frequency to $\hbar\omega_0 = 215$meV (1740 cm$^{-1}$), which corresponds to the C=O bond stretching frequency\cite{Shalabney2015, Kurouski2017a} and is also where surface plasmons in two-dimensional materials such as graphene typically reside\cite{Low2017}. Vibrational circular dichroism (VCD) is the conventional method of detecting molecule chirality in the mid-infrared frequency range\cite{Kurouski2017a}.

We briefly discuss material platforms that can support chiral surface plasmons before presenting the general theory. The most thoroughly studied material is twisted bilayer graphene (TBG). The chiral optical response of TBG was first observed using a far-field circular dichroism measurement\cite{Kim2016a}. Drude-like electrical conductivity and chiral conductivities were shown using theoretical linear response calculations\cite{Stauber2018a, Stauber2018, Lin2020}. It may also be possible to further engineer the chiral optical properties by stacking multiple layers with a constant relative twist between consecutive layers\cite{Kim2024a}. Alternatively, the theory presented in this work may also be extended to describe other polaritonic excitations, such as phonon polaritons in twisted stacks of 2D or thin film materials. For example, phonon polaritons in twisted bilayer $\alpha$-MoO$_3$ exhibit twist angle-dependent propagation\cite{Hu2020}, where the interlayer coupling between phonon polaritons can endow it with optical chirality\cite{Enders2025}. 

{\color{NavyBlue}\emph{Chiral surface plasmons}}.---We first outline a quantization procedure for the chiral surface plasmon\cite{Ferreira2020} and examine its properties. Within the Coulomb gauge ($\nabla\cdot\mathbf{A}=0$), the transverse electric and magnetic fields of the plasmon are given by
\begin{equation}
	\mathbf{E}_T(\mathbf{r},t)=-\frac{\partial \mathbf{A}(\mathbf{r},t)}{\partial t},\ \mathbf{B}(\mathbf{r},t)=\nabla\times\mathbf{A}(\mathbf{r},t).
\end{equation} 
Since all components of the plasmon field will be transverse, we will drop the $T$ subscript on the electric field from this point onwards. For fields in a lossless and weakly dispersive medium, the quantized vector potential may be written as a mode expansion\cite{Rivera2020a}
\begin{equation}
	\hat{\mathbf{A}}(\mathbf{r},t)=\sum_{\mathbf{q}\nu}\sqrt{\frac{\hbar}{2\epsilon_0 \omega_{\mathbf{q}\nu}SL_{\mathbf{q}\nu}}}\mathbf{A}_{\mathbf{q}\nu}(z)\hat{a}_{\mathbf{q}\nu}e^{i\mathbf{q}\cdot\mathbf{x}}e^{-i\omega_{\mathbf{q}\nu}t} + \mathrm{h.c.}.
\end{equation}
where $\hat{a}_{\mathbf{q}\nu}$ is the annihilation operator of a mode with in-plane momentum $\mathbf{q}$ and mode index $\nu$ at frequency $\omega_{\mathbf{q}\nu}$, $\mathbf{A}_{\mathbf{q}\nu}(z)$ is the polarization vector, $S$ is the surface area, and $L_{\mathbf{q}\nu}$ is an out-of-plane normalization length for the surface plasmon field. The normalization length $L_{\mathbf{q}\nu}$ is defined such that the energy contained in a single mode\cite{Tretyakov2003a} is equivalent to that of a harmonic oscillator with energy $\hbar\omega_{\mathbf{q}\nu}$ and may also be interpreted as the decay length of the surface plasmon field. The mode volume of the surface plasmon is given by $V_{\mathbf{q}\nu}=SL_{\mathbf{q}\nu}$.

The dispersion and fields of the chiral plasmon can be determined by solving the electromagnetic boundary conditions at the bianisotropic conducting surface. We assume that the electric conductivity is given by the {Drude model $\sigma_e=iD/\omega$}, which is well known to support a TM surface plasmon mode\cite{Jablan2009}. In addition, we will add a constant $\sigma_\chi$ that will cause a mixing of TM and TE field components, thus making the plasmon chiral. The boundary conditions then read\cite{Lin2020}
\begin{subequations}
\begin{gather}
		\hat{\mathbf{z}}\times\left(\mathbf{H}_+-\mathbf{H}_-\right) = \sigma_e\left(\frac{\mathbf{E}_++\mathbf{E}_-}{2}\right)-\sigma_\chi \left(\frac{\mathbf{H}_+ +\mathbf{H}_-}{2}\right) \\
		\hat{\mathbf{z}}\times\left(\mathbf{E}_+-\mathbf{E}_-\right) = -\sigma_\chi \left(\frac{\mathbf{E}_+ +\mathbf{E}_-}{2}\right)
\end{gather}	
\end{subequations}
where the subscript $+$($-$) refers to the fields above(below) the conducting surface at $z=0$. Assuming a vacuum dielectric environment, the dispersion relation for the chiral surface plasmons is given by
\begin{equation}
			\left(
		1-\frac{q'^2}{ k_0^2}\frac{\sigma_{\chi}^2}{4}+i\frac{q'}{2\epsilon_0 \omega}\sigma_{e}
		\right)
		\left(
		\frac{q'^2}{ k_0^2}-\frac{\sigma_{\chi}^2}{4}-i\frac{q'}{2\epsilon_0\omega}\sigma_{e}
		\right)
		=0
\end{equation}
where $q'=\sqrt{q^2- k_0^2}$ and $k_0=\omega/c=2\pi/\lambda_0$ is the free space wavevector of light. In the limit $\sigma_\chi=0$, the expression in the first bracket reduces to the dispersion relation of the TM surface plasmon while the expression in the second bracket becomes the dispersion relation of the TE plasmon. Here, we will be focusing on the surface plasmon given by the dispersion in the first bracket. A discussion of the second bracket dispersion is given in the supplementary information. 

\begin{figure}
    \centering
    \includegraphics{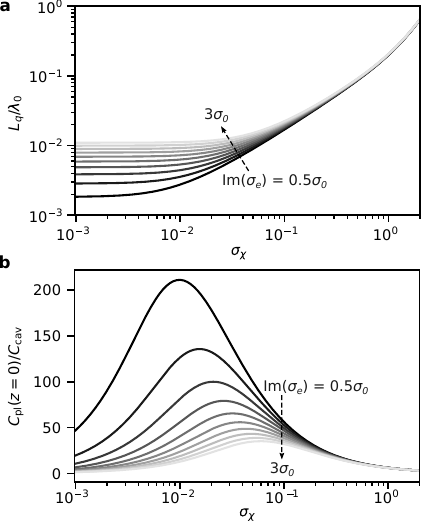}
    \caption{Properties of the chiral plasmon. \textbf{a} Dependence of the confinement, $L_q/\lambda_0$, on conductivities $\sigma_\chi$ and $\sigma_e$. The electric conductivity $\sigma_e$ is written in units of the universal conductivity $\sigma_0=e^2/4\hbar$. \textbf{b} Enhancement of optical chirality of the surface plasmon over that of a chiral cavity mode.}
    \label{fig:1-2}
\end{figure}

The polarization vector $\mathbf{A}_\mathbf{q}(z)$ for the chiral plasmon is found to be
\begin{equation}
	\mathbf{A}_\mathbf{q}(z)=
	\begin{cases}
		\left(
		\mathbf{u}_+^\mathrm{TM}+i\frac{\sigma_\chi}{2}\mathbf{u}_+^\mathrm{TE}
		\right)e^{-q'z}, & z>0 \\
		\left(
		\mathbf{u}_-^\mathrm{TM}-i\frac{\sigma_\chi}{2}\mathbf{u}_-^\mathrm{TE}
		\right)e^{q'z}, & z<0
	\end{cases}
\end{equation}
where $\mathbf{u}_\pm^\mathrm{TM}=i\hat{\mathbf{q}}\mp q/q' \hat{\mathbf{z}}$, $\mathbf{u}_\pm^\mathrm{TE}=\hat{\mathbf{q}}_\perp$, and $\hat{\mathbf{q}}_\perp$ is a unit vector perpendicular to $\mathbf{q}$ in the conducting plane. Note how $\sigma_\chi$ directly controls the degree of TM-TE hybridization for the surface plasmon mode. The normalization length $L_q$ is then given by
\begin{equation}
	L_q=\frac{1}{q'^3}\left(q^2+q'^2\right)+\frac{1}{q'}\frac{\sigma_\chi^2}{4}.
\end{equation}
The degree of confinement of the surface plasmon field with respect to a propagating wave at the same frequency $\omega$ is defined as $L_q/\lambda_0$. {Confinement as a function of $\sigma_\chi$ and $\sigma_e$ at a representative molecular vibrational frequency of $\hbar\omega = 215$meV (1740 cm$^{-1}$) is shown in \cref{fig:1-2}a.} Increasing $\sigma_\chi$ decreases the localization of the surface plasmon mode and thus increases the mode volume. Intuitively, this can be understood as a consequence of increasing the TE character of the mode. TE surface plasmon modes are weakly confined to the surface since its surface waves are bounded by the much weaker magnetic coupling. Hence, increasing $\sigma_\chi$ results in an increasing(decreasing) degree of TM-TE hybridization(field confinement). 

A quantity that can capture both the TM-TE hybridization and field confinement is the optical chirality, defined as $C=\frac{\epsilon_0\omega}{2}\mathrm{Im}(\langle\hat{\mathbf{E}}\cdot\hat{\mathbf{B}}^*\rangle)$\cite{Tang2010, Bliokh2011}, where the expectation value is taken with respect to the Fock state $|n\rangle$. For the surface plasmon, the optical chirality is 
\begin{equation}
    C_{\mathrm{pl}}(z)=\frac{\hbar\omega}{2V_q}\frac{q^2}{q'}\frac{\sigma_\chi}{2}e^{-2q'|z|}.
\end{equation}
Note that the optical chirality is uniform in the in-plane $(x,y)$ spatial dimensions. To clearly observe the effects of field confinement provided by the chiral surface plasmon, we will be be normalizing quantities obtained for the chiral surface plasmon with respect to identical quantities derived for a chiral mirror cavity. Consider a chiral cavity of length $L_{\mathrm{cav}}$ constructed from mirrors of surface area $S$ such that the total mode volume is $V_{\mathrm{cav}}=SL_{\mathrm{cav}}$. The electromagnetic fields in the cavity form standing waves with polarization vector $\mb{E}_{\mathrm{cav}}(z)=\cos(k_0z)\hat{\mb{x}}-\chi\sin(k_0z)\hat{\mb{y}}$ where $\chi=\pm$ is the handedness of the mode and the $z$ is the direction normal to the mirror surfaces\cite{Schafer2023}. The optical chirality of circularly polarized light confined to a chiral cavity is given by $C_{\mathrm{cav}}=\hbar\omega k_0/4V_{cav}$. We will be setting the cavity length to be the shortest cavity length that supports a cavity mode at frequency $\omega$, which is given by $L_{\mathrm{cav}}=\lambda_0/2=\pi c/\omega$. The enhancement of optical chirality in the surface plasmon over the cavity mode $C_{\mathrm{pl}}(z=0)/C_{\mathrm{cav}}$ is shown in \cref{fig:1-2}b. There exists an optimal value of $\sigma_\chi$ that maximizes the optical chirality enhancement due to a trade-off between TM-TE hybridization and field confinement. {The optimal value will depend on factors such as the carrier density and frequency.} The dependence on carrier density is captured in the calculations by varying $\sigma_e$. Increasing $\sigma_e$ (which corresponds to larger carrier densities) weakens the field confinement $L_q/\lambda_0$ and optical chirality $C$.

\begin{figure}
      \centering
      \includegraphics{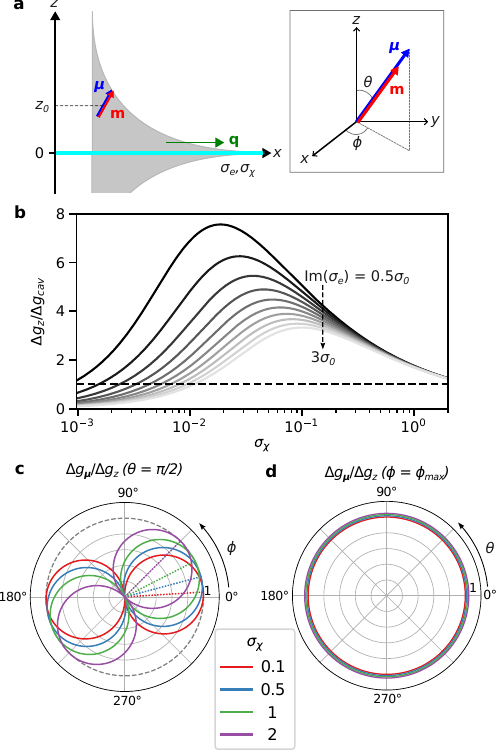}	
      \caption{Coupling of the chiral surface plasmon with a single CM. \textbf{a} Schematic representation of the coupled system. The surface plasmon is confined to the chiral surface at $z=0$ with in-plane momentum $\mathbf{q}$ along the $x$-axis. The CM is represented as a parallel electric($\boldsymbol{\mu}$) and magnetic($\mathbf{m}$) transition dipole moment at a height $z_0$ above the chiral surface. The orientation of the transition dipole moment is represented using the two polar angles $\theta,\phi$ as shown in the inset. \textbf{b} Enhancement of the chiral discrimination factor $\Delta g$ for a chiral surface plasmon over that of a chiral cavity.  \textbf{c} Dependence of chiral discrimination factor as a function of $\phi$ with the in-plane angle fixed to $\theta=\pi/2$. The colored dashed lines indicate the $\phi_{\textrm{max}}$ angle at which the discrimination factor is equivalent to $\Delta g_z$. \textbf{d} Chiral discrimination factor as a function of $\theta$ with $\phi=\phi_{\textrm{max}}$ fixed. Discrimination factors in \textbf{c,d} are normalized to $\Delta g_z$, the discrimination factor for a $z$-oriented dipole.}
      \label{fig:2}
\end{figure}

\emph{\color{NavyBlue}Coupling strength with chiral molecules.}---Consider the interaction between a single CM at position $\mathbf{R}_0=(\mathbf{x}_0,z_0)$ and the surface plasmon as shown in \cref{fig:2}a. The CM can be represented with parallel electric and magnetic transition dipole moments\cite{Mun2020, Schafer2023}. Assuming a two-level system, these dipole operators may then be written as $	\hat{\boldsymbol{\mu}}=\boldsymbol{\mu}_{12}(\hat{\sigma}^\dagger+\hat{\sigma}),\ \hat{\mathbf{m}}=ic\xi\boldsymbol{\mu}_{12}(\hat{\sigma}^\dagger -\hat{\sigma})$ where $\boldsymbol{\mu}_{12}$ is the electric transition dipole moment, $c$ is the speed of light, and $\xi$ represents the chirality of the CM. For the polar angles defined in the inset of \cref{fig:2}a, the transition dipole moment can be written as $\boldsymbol{\mu}_{12}=\mu_{12}(\sin\theta\cos\phi,\sin\theta\sin\phi,\cos\theta)$. When $\xi=\pm1$, the CM is perfectly left(right) handed. The interaction Hamiltonian for the surface plasmon and a single CM at position $\mathbf{R}_0=(\mathbf{x}_0,z_0)$ is then
\begin{equation}
	\hat{H}_I=-\hat{\boldsymbol{\mu}}\cdot\hat{\mathbf{E}}-\hat{\mathbf{m}}\cdot\hat{\mathbf{B}}=-\sum_\mathbf{q} g_{\boldsymbol{\mu}}(\mathbf{q},z_0)\hat{a}_\mathbf{q}\hat{\sigma}^\dagger e^{i\mathbf{q}\cdot\mathbf{x}_0}+\mathrm{h.c.}
\end{equation}
where the coupling strength $g_{\boldsymbol{\mu}}(\mathbf{q},z_0)$ is
\begin{align}\label{eq:g}
	g_{\boldsymbol{\mu}}(\mathbf{q};z_0)=\sqrt{\frac{\hbar}{2\epsilon_0\omega_\mathbf{q}V_q}}&\boldsymbol{\mu}_{12}\cdot\Bigl[\mb{E}_\mb{q}(z_0)+ic\xi\mb{B}_\mb{q}(z_0)
\Bigr]
\end{align}
and $\mb{E}_\mb{q}(z),\mb{B}_\mb{q}(z)$ are the polarization vectors of the electric and magnetic fields. Expressions for the polarization vectors are given in the supplementary information. When the transition dipole moment $\boldsymbol{\mu}_{12}$ is oriented along the $z$ axis, the coupling strength becomes
\begin{equation}
	g_z(\mathbf{q};z_0)=i\sqrt{\frac{\hbar\omega_\mathbf{q}}{2\epsilon_0V_q}}\mu_{12}\frac{q}{q'}\left(1+\frac{q'}{k_0}\frac{\xi\sigma_\chi}{2}\right)e^{-q'|z_0|}.
\end{equation}
As expected, the handedness of the CM and surface plasmon dictate the coupling strength. The coupling strength is stronger(weaker) when the CM and surface plasmon have equal(opposite) handedness.

The chiral discrimination factor $\Delta g$ may be defined as the difference in coupling strength of a CM with left- and right-handed light. For chiral surface plasmons coupled to a $z$-oriented CM, the discrimination factor is $\Delta g_z=\sqrt{\frac{2\hbar\omega_q}{\epsilon_0V_q}}\mu_{12}\xi\frac{q}{k_0}\frac{|\sigma_\chi|}{2}e^{-q'|z_0|}$. For chiral cavity modes with the dipole aligned with the electric field of the mode, the discrimination factor is $\Delta g_{\mathrm{cav}}=\sqrt{\frac{2\hbar\omega}{\epsilon_0 V_{\mathrm{cav}}}}\mu_{12}\xi$. Once again, assuming that the cavity length is $L_{\mathrm{cav}}=\lambda_0/2$, the enhancement of the discrimination factor provided by the surface plasmon is 
\begin{equation}
	\frac{\Delta g_z}{\Delta g_{\mathrm{cav}}}=\sqrt{\frac{\pi}{k_0 L_q}}\frac{q}{k_0}\frac{|\sigma_\chi|}{2}e^{-q'|z_0|}.
\end{equation}
This definition of the enhancement factor enables us to examine the effect of the plasmon confinement and chirality on the chiral discrimination factor independent of the molecule parameters. The enhancement as a function of $\sigma_\chi$ for a $z$-oriented dipole placed at $z_0=1$nm is shown in \cref{fig:2}b. For the chiral cavity, the moment is aligned with the polarization vector to give maximal coupling strength. Similar to the enhancement of optical chirality, the enhancement of the chiral discrimination factor is strongest when the degree of TM-TE hybridization and field localization is optimally balanced. 

In general, the discrimination factor $\Delta g_{\boldsymbol{\mu}}$ can be calculated for an arbitrarily oriented CM using \cref{eq:g}. The polar angles that define the CM orientation are shown in the inset of \cref{fig:2}a. When the CM is parallel to the conducting surface ($\theta=\pi/2$), the discrimination factor has a strong dependence on the azimuthal angle $\phi$. There exists a unique in-plane orientation, $\phi_{\textrm{max}}$,  at which an in-plane dipole couples as strongly as a perpendicular (z-oriented) dipole. When $\phi=\phi_\mathrm{max}\pm\pi/2$, the CM decouples from the surface plasmon field. The full dependence of the discrimination factor on $\phi$ at $\theta=\pi/2$ is shown in \cref{fig:2}c. Interestingly, when $\phi=\phi_\mathrm{max}$, the discrimination factor is identical for all values of $\theta$ at any given value of $\sigma_\chi$ (see \cref{fig:2}d). Hence, the coupling strength and discrimination factor are determined by the projection of the transition dipole moment onto a two-dimensional plane with normal vector $\hat{\mathbf{n}}=(-\sin(\phi_\mathrm{max}), \cos(\phi_\mathrm{max}), 0)$. In contrast, the coupling strength of a dipole with chiral cavity modes (or any type of free-space propagating light) is dependent on the projection of the dipole moment onto the one-dimensional polarization vector. This geometric difference produces a $\sqrt{2}$ orientation-averaged boost in chiral discrimination for plasmonic platforms.
\begin{figure}
      \centering
      \includegraphics{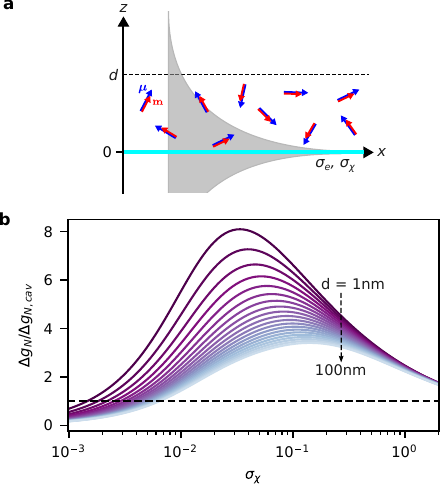}	
      \caption{Coupling of the chiral surface plasmon with an ensemble of $N$ CMs. \textbf{a} The $N$ CMs are uniformly distributed within a thin layer of thickness $d$ above the chiral surface. Random orientation is assumed. \textbf{b} Enhancement of discrimination factor for the $N$ CM ensemble. An identical randomly oriented ensemble distributed over thickness $d$ is assumed when calculating the discrimination factor for the cavity $\Delta g_{N,\mathrm{cav}}$. The enhancement is plotted as a function of $\sigma_\chi$ for thicknesses $d$ in the range of 1nm to 100nm and the electric conductivity is fixed to $\sigma_e=i\sigma_0$. } 
      \label{fig:3}
\end{figure}

Now consider $N$ CMs uniformly distributed with random orientations in a layer of thickness $d$ above the conducting surface as shown in \cref{fig:3}a. Following the treatment of \cite{Gonzalez-Tudela2013}, we assume that there are $N_L$ layers of CMs between $z=0$ and $z=d$ with $N_S$ in each layer distributed equally over area $S$ such that the total number of CMs is $N=N_L N_S$ and the CM density is $n=N/(dS)$. Building up a collective operator $\hat{D}_\mathbf{q}$ of the $N$ CMs, we may write the interaction Hamiltonian as 
\begin{equation}
	\hat{H}_I = -\sum_\mathbf{q} g_N(\mathbf{q})\hat{a}_\mathbf{q}\hat{D}^\dagger_{\mathbf{q}}+\mathrm{h.c.}
\end{equation}
where the effective coupling strength to the $N$ emitters is given by 
\begin{equation}
	g_N(\mathbf{q})=\sqrt{N_S\sum_{j=1}^{N_L}\langle |g_{\boldsymbol{\mu}}(\mathbf{q},z_j)|^2\rangle}=\sqrt{n\int_0^d S\langle|g_{\boldsymbol{\mu}}(\mathbf{q},z)|^2\rangle dz}
\end{equation}
and $\langle |g_{\boldsymbol{\mu}}|^2\rangle$ refers to the rotationally averaged single-emitter coupling strength. If we assume a molecule dipole moment of 1 Debye\cite{Shalabney2015} and a plasmon damping rate of 10 cm$^{-1}$\cite{Basov2018}, the coupled system is found to enter the strong coupling regime at a molecular density of 10$^{20}$cm$^{-3}$ for a 100nm-thick film. The enhancement in discrimination factor for the $N$ CM case provided by the surface plasmons over the cavity modes is shown in \cref{fig:3}b as a function of both $\sigma_\chi$ and thickness $d$. As we have observed for the single-emitter case, the coupling strength has an optimal value of $\sigma_\chi$ due to a trade-off between TM-TE hybridization and field localization. The enhancement also becomes stronger as the CMs are distributed over a thinner region above the conducting surface. This is a consequence of the exponential localization of the surface plasmon field. While the coupling strength to a surface plasmon exponentially diminishes as a function of the distance $z$ to the conducting surface, the coupling strength to a chiral cavity mode is independent of the emitter's position. It is, therefore, expected that enhancement provided by the surface plasmon becomes weaker as emitters are placed within the exponentially decaying tail of the field.

\begin{figure}
      \centering
      \includegraphics{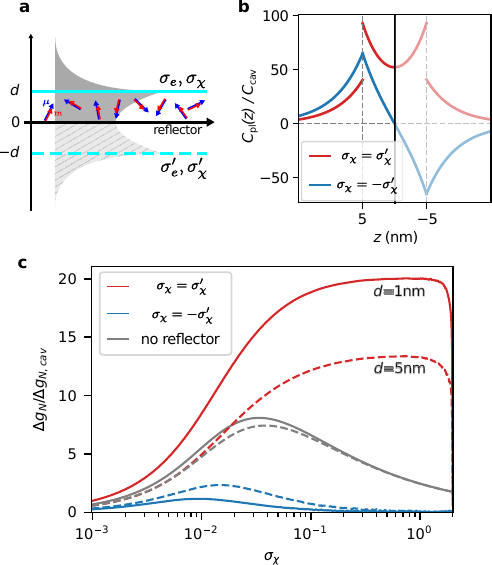}	
      \caption{Chiral surface plasmons with a reflector. \textbf{a} Schematic representation of the chiral surface with a reflector. The reflector is placed at $z=0$ with the chiral surface at $z=d$. Image charges are formed at $z=-d$ and are represented as $\sigma_e',\sigma_\chi'$. The relation between $\sigma_\chi$ and $\sigma_\chi'$ will depend on the nature of the reflector. An ensemble of randomly oriented CMs are distributed between $z=0$ and $z=d$. \textbf{b} Optical chirality of the surface plasmon field in the presence of a handedness-preserving ($\sigma_\chi=\sigma_\chi'$) and normal ($\sigma_\chi=-\sigma_\chi'$) reflector. \textbf{c} Enhancement of discrimination factor for the $N$ CM ensemble with reflectors. The red(blue) curves correspond to the handedness-preserving(normal) reflectors. The grey curve is the enhancement for the chiral surface without a reflector. The solid(dashed) curves correspond to $d=1$nm($d=5$nm).}
      \label{fig:4}
\end{figure}

\emph{\color{NavyBlue}Enhanced field localization with chiral reflectors}.---Because the chiral discrimination factor scales with local field confinement, we consider the classic strategy of placing a metallic reflector beneath the chiral conducting sheet\cite{Alonso-Gonzalez2017, Lee2019}. It should be noted that the CMs exist in the spacer layer between the chiral conducting surface and the reflector\cite{Chrisey2003}.
The effect of a reflector will be to form image charges of the surface plasmon excitation on the opposite side of the reflector (see \cref{fig:4}a). The reflector is placed at $z=0$ with the chiral conducting surface at $z=d$ and image charges formed at $z=-d$. We thus model the effect of a reflector by considering the acoustic chiral plasmons of a bilayer system with a layer separation of $2d$ with the CMs placed between $z=0$ and $d$. Details of the bilayer calculation are presented in the supplementary information.

If the reflector preserves handedness (as in the case of the chiral mirrors used for chiral cavities), the image charges will have the same handedness as the original surface plasmon\cite{Plum2015, Voronin2022a}. The image charges will have opposite handedness for a reflector that does not preserve handedness (e.g., a gold metal reflector). The optical chirality of the fields for both handedness-preserving and non-preserving reflectors is shown in \cref{fig:4}b. The handedness-preserving reflector gives a surface plasmon field with uniformly enhanced optical chirality throughout the region between the conducting surface and the reflector. In contrast, the reflector that does not preserve handedness results in a weaker optical chirality in regions close to the reflector. 

\Cref{fig:4}c quantifies how each mirror modifies the ensemble-averaged discrimination factor for randomly oriented CMs dispersed between the two plates.  When compared with the surface plasmon without a reflector, we find that the handedness-preserving reflector provides a significant improvement in the chiral discrimination factor. The enhancement in discrimination factor extends up to large values of $\sigma_\chi$. For the chiral surface without a reflector, large values of $\sigma_\chi$ weakened the confinement of the field, thus resulting in a smaller discrimination factor. When a handedness-preserving reflector is introduced, the strong field confinement is enforced by the reflecting boundary and is maintained even for large values of $\sigma_\chi$. A reflector that does not preserve handedness reduces the chiral discrimination factor due to the weak optical chirality of the fields. 

\emph{\color{NavyBlue}Discussion}.---In this work, we presented a general theory that describes the interaction between chiral surface plasmons supported by twisted atomic bilayers and chiral molecules described within the dipole approximation. Due to the strong field confinement provided by the surface plasmons, we found an enhanced chiral discrimination factor compared to chiral mirror cavities for both single molecules and $N$ molecule ensembles. Finally, we presented a setup incorporating chiral mirror reflectors to further enhance the confinement and thus also the chiral discrimination factor.

As noted in the introduction, the spatially uniform optical chirality is one of the main advantages of utilizing chiral surface plasmons supported by twisted atomic bilayers. Most chiral plasmons in geometrically chiral nanostructures made of non-chiral materials (e.g., gold) exhibit optical chirality that varies spatially and even changes sign\cite{Schaferling2012}. 

One possible route to further enhancing the light-matter interaction and chiral discrimination factor would be to incorporate non-chiral patterning, such as ribbons or disks, into the twisted atomic bilayers. Such nanostructures would enable further confinement of the surface plasmon field along the in-plane spatial dimensions without changing the sign of optical chirality. Indeed, non-chiral surface plasmons in patterned nanoribbon or nanodisk structures have been shown to facilitate light-matter interactions\cite{Koppens2011, Rodrigo2015}. In principle, the confinement ratio can reach up to $(L_q/\lambda_0)^3\sim 10^{-9}$ when the plasmon field is confined along all three spatial dimensions. The chiral discrimination factor can be expected to be even larger than the values reported in this paper under full confinement in all spatial dimensions. Finally, the combination of chiral geometric structures and intrinsically chiral materials may also provide a promising route to enhancing chiral light-matter interactions\cite{Huang2022}.

\emph{\color{NavyBlue}Acknowledgements.}---This work was supported the Multi-University Research Initiative (MURI) on Twist-Optics, sponsored by the Office of Naval Research under Grant No. N00014-23-1-2567.

\bibliography{references}

\end{document}


\title{Supplementary Information: Enhanced enantiomer discrimination with chiral surface plasmons}

\author{Sang Hyun Park}
\affiliation{Department of Electrical \& Computer Engineering, University of Minnesota, Minneapolis, MN, USA}
\affiliation{Department of Physics, The University of Texas at Austin, Austin, TX, USA}
\author{Phaedon Avouris}
\affiliation{Department of Electrical \& Computer Engineering, University of Minnesota, Minneapolis, MN, USA}
\author{Jennifer A. Dionne}
\affiliation{Department of Materials Science and Engineering, Stanford University School of Engineering, Stanford, CA, USA}
\affiliation{Department of Radiology, Stanford University School of Medicine, Stanford, CA, USA}
\author{Joshua D. Caldwell}
\affiliation{Interdisciplinary Materials Science Graduate Program, Vanderbilt University, Nashville, TN, USA}
\affiliation{Vanderbilt University, Department of Mechanical Engineering, Nashville, TN, USA}
\author{Tony Low}
\email{tlow@umn.edu}
\affiliation{Department of Electrical \& Computer Engineering, University of Minnesota, Minneapolis, MN, USA}

\maketitle

\section{Quantization of the chiral surface plasmon}
In this section we outline the macroscopic quantization procedure of a chiral surface plasmon supported by a lossless and dispersive conducting surface\cite{Ferreira2020}. In the Coulomb gauge ($\nabla\cdot\mathbf{A}=0$), the transverse electric and magnetic fields are written in terms of the vector potential as
\begin{equation}
	\mathbf{E}_T(\mathbf{r},t)=-\frac{\partial \mathbf{A}(\mathbf{r},t)}{\partial t},\quad \mathbf{B}(\mathbf{r},t)=\nabla\times\mathbf{A}(\mathbf{r},t).
\end{equation}  
Since all the plasmon fields will be transverse, we will drop the $T$ subscript from this point forward. Further assuming translational invariance of the conducting surface and its dielectric environment, the vector potential may be written as a mode expansion
\begin{equation}
	\mathbf{A}(\mathbf{r},t)=\sum_{\mathbf{q}\nu}\frac{1}{\sqrt{S}}e^{i\mathbf{q}\cdot\mathbf{x}}e^{-i\omega_{\mathbf{q}\nu}t}\mathbf{A}_{\mathbf{q}\nu}(z)\alpha_{\mathbf{q}\nu}+\mathrm{h.c.}
\end{equation}
where $\mathbf{q}$ is the in-plane momentum, $\mathbf{x}$ is the in-plane spatial coordinate, $\nu$ is a mode index, $\mathbf{A}_{\mathbf{q}\nu}(z)$ is the mode polarization vector, and $\alpha_{\mathbf{q}\nu}$ is the mode amplitude. 

We now normalize the electromagnetic energy in each mode such that it is equal to $\hbar\omega_{\mathbf{q}\nu}$, the energy of a harmonic oscillator. The macroscopic material properties of a chiral medium are given by $\epsilon(\omega,z),\ \mu(\omega,z),\ \kappa(\omega,z)$ which define the constitutive relations
\begin{equation}\label{eq:constitutive}
	\begin{pmatrix}
		\mathbf D \\ \mathbf B
	\end{pmatrix}
	=\begin{pmatrix}
		\epsilon_0\epsilon(\omega) & i\kappa(\omega)/c_0 \\ -i\kappa(\omega)/c_0 & \mu_0\mu(\omega)
	\end{pmatrix}
	\begin{pmatrix}
		\mathbf E \\ \mathbf H
	\end{pmatrix}.
\end{equation}
Using the mode expansion for the electric and magnetic fields, the electromagnetic energy of a mode at $\omega_{\mathbf{q}\nu}$ in a dispersive chiral media is\cite{Tretyakov2003a}
\begin{align}
	U_{\mathbf{q}\nu}&=\iiint d^3r \frac{|\alpha_{\mathbf{q}\lambda}|^2}{S}\Bigg[
	\epsilon_0\frac{\partial(\omega\epsilon)}{\partial \omega}|\mathbf{E}_{\mathbf{q}\nu}(z)|^2+\mu_0\frac{\partial(\omega\mu)}{\partial \omega}|\mathbf{H}_{\mathbf{q}\nu}(z)|^2 +\frac{2}{c_0}\frac{\partial(\omega\kappa)}{\partial \omega}\mathrm{Im}(\mathbf{E}^*_{\mathbf{q}\nu}(z)\cdot\mathbf{H}_{\mathbf{q}\nu}(z))
	\Bigg] \nonumber \\
	&= \frac{1}{2}\left(\alpha_{\mathbf{q}\nu}^*\alpha_{\mathbf{q}\nu}+\alpha_{\mathbf{q}\nu}\alpha_{\mathbf{q}\nu}^*\right)\int dz \Biggl[
	\epsilon_0\frac{\partial(\omega\epsilon)}{\partial \omega}|\mathbf{E}_{\mathbf{q}\nu}(z)|^2+\mu_0\frac{\partial(\omega\mu)}{\partial \omega}|\mathbf{H}_{\mathbf{q}\nu}(z)|^2 +\frac{2}{c_0}\frac{\partial(\omega\kappa)}{\partial \omega}\mathrm{Im}(\mathbf{E}^*_{\mathbf{q}\nu}(z)\cdot\mathbf{H}_{\mathbf{q}\nu}(z))
	\Biggr] \label{eq:em_energy}
\end{align} 
where $\mathbf{E}_{\mathbf{q}\nu}(z),\mathbf{H}_{\mathbf{q}\nu}(z)$ are the polarization vectors for the electric and magnetic fields. For the second equality, we have intergrated over the in-plane spatial dimensions.

Following the canonical quantization process, we elevate the mode amplitudes into creation and annihilation operators by \cite{Loudon2000}
\begin{equation}
	\alpha_{\mathbf{q}\nu}\rightarrow \sqrt{\frac{\hbar}{2\epsilon_0 L_{\mathbf{q}\nu}\omega_{\mathbf{q}\nu}}} \hat{a}_{\mathbf{q}\nu},\quad \alpha_{\mathbf{q}\nu}^*\rightarrow \sqrt{\frac{\hbar}{2\epsilon_0 L_{\mathbf{q}\nu}\omega_{\mathbf{q}\nu}}} \hat{a}^\dagger_{\mathbf{q}\nu}
\end{equation}
where $L_{\mathbf{q}\nu}$ is the normalization length in the $z$-direction. Requiring that the energy of each mode is equal to $\hbar\omega_{\mathbf{q}\nu}$ fixes the normalization length to 
\begin{align}\label{eq:normalization}
L_{\mathbf{q}\nu}=\frac{1}{2\epsilon_0\omega_{\mathbf{q}\nu}^2}	\int dz \Biggl[
	\epsilon_0\frac{\partial(\omega\epsilon)}{\partial \omega}|\mathbf{E}_{\mathbf{q}\nu}(z)|^2+\mu_0&\frac{\partial(\omega\mu)}{\partial \omega}|\mathbf{H}_{\mathbf{q}\nu}(z)|^2 +\frac{2}{c_0}\frac{\partial(\omega\kappa)}{\partial \omega}\mathrm{Im}(\mathbf{E}^*_{\mathbf{q}\nu}(z)\cdot\mathbf{H}_{\mathbf{q}\nu}(z))
	\Biggr].
\end{align}
The quantized vector potential is then
\begin{equation}
	\hat{\mathbf{A}}(\mathbf{r},t)=\sum_{\mathbf{q}\nu}\sqrt{\frac{\hbar}{2\epsilon_0 SL_{\mathbf{q}\nu}\omega_{\mathbf{q}\nu}}}e^{i\mathbf{q}\cdot\mathbf{x}}e^{-i\omega_{\mathbf{q}\nu}t}\mathbf{A}_{\mathbf{q}\nu}(z)\hat{a}_{\mathbf{q}\nu}+\mathrm{h.c.}
\end{equation}
where we can identify the mode volume as $V_{\mathbf{q}\nu}=SL_{\mathbf{q}\nu}$. 

\section{Dispersion relation and field profile of the chiral surface plasmon}
The polarization vectors are determined by applying boundary conditions at the conducting surface. Consider a conducting surface in the $z=0$ plane with both electric $\sigma_e$ and chiral $\sigma_\chi$ conductivities encapsulated in a dielectric with constant permittivity $\epsilon_s=1$. The polarization vector for the vector potential may be written as 
\begin{equation}
	\mathbf{A}_\mathbf{q}(z)=
	\begin{dcases}
		\left(
		A_+^\mathrm{TM}\mathbf{u}_+^\mathrm{TM}+A_+^\mathrm{TE}\mathbf{u}_+^\mathrm{TE}
		\right)e^{-q'z}, & z>0 \\[0.5em]
		\left(
		A_-^\mathrm{TM}\mathbf{u}_-^\mathrm{TM}+A_-^\mathrm{TE}\mathbf{u}_-^\mathrm{TE}
		\right)e^{q'z}, & z<0
	\end{dcases}
\end{equation}
where $\mathbf{u}_\pm ^\mathrm{TM}=i\hat{\mathbf{q}} \mp q/q'\hat{\mathbf{z}}$, $\mathbf{u}_{\pm}^{\mathrm{TE}}=\hat{\mathbf{q}}_\perp$, $q'=\sqrt{q^2-\omega^2/c^2}$, and $c=1/\sqrt{\epsilon_0\mu_0}$. The mode index is dropped because only a single mode exists at each wavevector. The coefficients $A_\pm^{\mathrm{TM/TE}}$ are determined by applying the boundary conditions at $z=0$ to the electric and magnetic fields. The polarization vectors for the electric and magnetic fields are explicitly given as
\begin{equation}
	\mathbf{E}_{\mathbf{q},\pm}(z)=i\omega
	\begin{pmatrix}
		iA^{\tm}_\pm \\[0.5em] A^\te_\pm \\[0.5em] \mp \dfrac{q}{q'}A^\tm_\pm
	\end{pmatrix}
	e^{\mp q'z},\quad
	\mb{B}_{\mb{q},\pm}(z)=
	\begin{pmatrix}
		\pm q'A^\te_\pm \\[0.5em] 
		\pm i \dfrac{\omega^2}{q'c^2}A^\tm_\pm \\[1em]
		iq A^\te_\pm
	\end{pmatrix}
	e^{\mp q'z}
\end{equation}
where the column vectors are written in the basis $(\hat{\mb{q}},\hat{\mb{q}}_\perp,\hat{\mb{z}})$. The boundary conditions at $z=0$ are 
\begin{gather}
		\hat{\mathbf{z}}\times\left(\mathbf{H}_+-\mathbf{H}_-\right) = \sigma_e\left(\frac{\mathbf{E}_++\mathbf{E}_-}{2}\right)-\sigma_\chi \left(\frac{\mathbf{H}_+ +\mathbf{H}_-}{2}\right) \\
		\hat{\mathbf{z}}\times\left(\mathbf{E}_+-\mathbf{E}_-\right) = -\sigma_\chi \left(\frac{\mathbf{E}_+ +\mathbf{E}_-}{2}\right).
\end{gather}	
The resulting set of equations may be written as a matrix equation
\begin{equation}\label{eq:matrix_dispersion}
	\begin{pmatrix}
		-i\dfrac{\sigma_\chi}{2} & -i\dfrac{\sigma_\chi}{2} & -1 & 1 \\
		i & -i & -\dfrac{\sigma_\chi}{2} & -\dfrac{\sigma_\chi}{2} \\
		-\mu_0\omega\dfrac{\sigma_e}{2}+i\dfrac{\omega^2}{q'c^2} & -\mu_0\omega\dfrac{\sigma_e}{2}+i\dfrac{\omega^2}{q'c^2} & q'\dfrac{\sigma_\chi}{2} & -q'\dfrac{\sigma_\chi}{2} \\
		i\dfrac{\omega^2}{q' c^2}\dfrac{\sigma_\chi}{2} & -i\dfrac{\omega^2}{q' c^2}\dfrac{\sigma_\chi}{2} & -q'+i\mu_0\omega\dfrac{\sigma_e}{2} & -q'+i\mu_0\omega\dfrac{\sigma_e}{2}
	\end{pmatrix}
	\begin{pmatrix}
		A_-^\mathrm{TM} \\[1em] A_+^\mathrm{TM} \\[1em] A_-^\mathrm{TE} \\[1em] A_+^\mathrm{TE}
	\end{pmatrix}
	=0.
\end{equation}
The dispersion relation is then given by setting the determinant of the matrix in \cref{eq:matrix_dispersion} to zero which gives
\begin{equation}\label{eq:dispersion}
			\left(
		1-\frac{q'^2}{\epsilon k_0^2}\frac{\sigma_{\chi}^2}{4}+i\frac{q'}{2\epsilon_0\epsilon \omega}\sigma_{e}
		\right)
		\left(
		\frac{q'^2}{\epsilon k_0^2}-\frac{\sigma_{\chi}^2}{4}-i\frac{q'}{2\epsilon_0\epsilon\omega}\sigma_{e}
		\right)
		=0.
\end{equation}

The eigenvector corresponding to the solution given by the first term in \cref{eq:dispersion} is
\begin{equation}
	(A_-^\mathrm{TM},\ A_+^\mathrm{TM},\ A_-^\mathrm{TE},\ A_+^\mathrm{TE})=(1,\ 1,\ -i\frac{\sigma_\chi}{2},\ i\frac{\sigma_\chi}{2}).
\end{equation}
The polarization vector for the vector poential is therefore
\begin{equation}
	\mathbf{A}_\mathbf{q}(z)=
	\begin{dcases}
		\left(
		\mathbf{u}_+^\mathrm{TM}+i\frac{\sigma_\chi}{2}\mathbf{u}_+^\mathrm{TE}
		\right)e^{-q'z}, & z>0 \\[0.5em]
		\left(
		\mathbf{u}_-^\mathrm{TM}-i\frac{\sigma_\chi}{2}\mathbf{u}_-^\mathrm{TE}
		\right)e^{q'z}, & z<0.
	\end{dcases}
\end{equation}
As noted in the main text, this solution becomes the TM plasmon mode in the limit $\sigma_\chi\rightarrow 0$. 
The eigenvector for the mode given by the second term in \cref{eq:dispersion} is 
\begin{equation}
	(A_-^\mathrm{TM},\ A_+^\mathrm{TM},\ A_-^\mathrm{TE},\ A_+^\mathrm{TE})=(-i\frac{\sigma_\chi}{2},\ i\frac{\sigma_\chi}{2},\ 1,\ 1)
\end{equation}
and becomes the TE plasmon mode in the limit $\sigma_\chi\rightarrow 0$. The TE plasmon requires $\mathrm{Im}\sigma_e<0$ and is not supported by a Drude-like conductivity but has been predicted to exist in graphene with interband transitions \cite{Mikhailov2007}. 

In this paper, we will be focusing on the TM-like mode. The electric and magnetic field polarization vectors of the TM-like mode are given by
\begin{equation}
		\mathbf{E}_{\mathbf{q},\pm}(z)=i\omega
	\begin{pmatrix}
		i \\[0.5em] \pm i\dfrac{\sigma_\chi}{2} \\[01em] \mp \dfrac{q}{q'}
	\end{pmatrix}
	e^{\mp q'z},\quad
	\mb{B}_{\mb{q},\pm}(z)=
	\begin{pmatrix}
		i q'\dfrac{\sigma_\chi}{2} \\[1em] 
		\pm i \dfrac{\omega^2}{q'c^2} \\[1em]
		\mp q \dfrac{\sigma_\chi}{2}
	\end{pmatrix}
	e^{\mp q'z}.
\end{equation}
The quantized electric and magnetic field operators are then
\begin{gather}
	\hat{\mathbf{E}}(\mathbf{r},t)=\sum_{\mathbf{q}}\sqrt{\frac{\hbar}{2\epsilon_0 V_{\mathbf{q}}\omega_{\mathbf{q}}}}e^{i\mathbf{q}\cdot\mathbf{x}}e^{-i\omega_{\mathbf{q}\nu}t}\mathbf{E}_{\mathbf{q}}(z)\hat{a}_{\mathbf{q}}+\mathrm{h.c.} \\
	\hat{\mathbf{B}}(\mathbf{r},t)=\sum_{\mathbf{q}}\sqrt{\frac{\hbar}{2\epsilon_0 V_{\mathbf{q}}\omega_{\mathbf{q}}}}e^{i\mathbf{q}\cdot\mathbf{x}}e^{-i\omega_{\mathbf{q}\nu}t}\mathbf{B}_{\mathbf{q}}(z)\hat{a}_{\mathbf{q}}+\mathrm{h.c.}
\end{gather}
The optical chirality for the Fock state $|n\rangle$ of the plasmon mode $\omega_\mathbf{q}$ is then calculated to be \cite{Schafer2023}
\begin{equation}
	C_\pm=\frac{\epsilon_0\omega}{2}\mathrm{Im}\left(\langle n| \hat{\mb{E}_\pm}\cdot\hat{\mb{B}}^\dagger_\pm|n\rangle\right)=\frac{\hbar\omega_\mathbf{q}}{2V_\mathbf{q}}\frac{q^2}{q'}\frac{\sigma_\chi}{2}e^{-2q'|z|}.
\end{equation}

To find the normalization length, we first define the $z$ dependent material parameters $\epsilon$ and $\kappa$. Including the conducting surface at $z=0$, we find \cite{Ferreira2020}
\begin{equation}
	\epsilon(\omega,z)=1+i\frac{\sigma_e}{\omega\epsilon_0}\delta(z),\quad \kappa(\omega,z)=\frac{\sigma_\chi c}{\omega}\delta(z).
\end{equation}
For $\sigma_e=iD/\omega$ and assuming a constant $\sigma_\chi$, we use the expression \cref{eq:normalization} to find
\begin{equation}
	L_q=\frac{1}{q'^3}\left(q^2+q'^2\right)+\frac{1}{q'}\frac{\sigma_\chi^2}{4}.
\end{equation}

\section{Bilayer conducting sheet calculation}
\begin{figure}[h]
	\centering
	\includegraphics{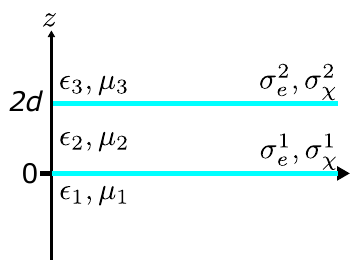}
	\caption{Setup for chiral surface plasmon calculation of a bilayer of conducting sheets}
	\label{fig:bilayersetup}
\end{figure}
Consider the bilayer chiral conducting sheet setup as shown in \cref{fig:bilayersetup}. The polarization vector of the vector potential is in general given as
\begin{equation}
	\mathbf{A}_{\mathbf{q}\nu}(z)=
	\begin{dcases}
		\left(
		A_{3,\nu}^\mathrm{TM}\mathbf{u}_3^{\mathrm{TM}} + A_{3,\nu}^\mathrm{TE}\mathbf{u}_3^{\mathrm{TE}}
		\right)e^{-q_3'z}, & z>2d \\[0.5em]
		\left(
		A_{2+,\nu}^\mathrm{TM}\mathbf{u}_{2+}^{\mathrm{TM}} + A_{2+,\nu}^\mathrm{TE}\mathbf{u}_{2+}^{\mathrm{TE}}
		\right)e^{q_2'z} + 
		\left(
		A_{2-,\nu}^\mathrm{TM}\mathbf{u}_{2-}^{\mathrm{TM}} + A_{2-,\nu}^\mathrm{TE}\mathbf{u}_{2-}^{\mathrm{TE}}
		\right)e^{-q_2'z} 
		, & 0<z<2d \\[0.5em]
		\left(
		A_{1,\nu}^\mathrm{TM}\mathbf{u}_1^{\mathrm{TM}} + A_{1,\nu}^\mathrm{TE}\mathbf{u}_1^{\mathrm{TE}}
		\right)e^{q_1'z}, & z<0 
	\end{dcases}
\end{equation}
where $\mathbf{u}_1^\mathrm{TM}=i\hat{\mathbf{q}}+q/q_1'\hat{\mathbf{z}}$, $\mathbf{u}_{2\pm}^\mathrm{TM}=i\hat{\mathbf{q}}\pm q/q_{2}'\hat{\mathbf{z}}$, $\mathbf{u}_3^\mathrm{TM}=i\hat{\mathbf{q}}-q/q_3'\hat{\mathbf{z}}$, and $\hat{\mathbf{u}}_j^\mathrm{TE}=\hat{\mathbf{q}}_\perp$. Applying boundary conditions at $z=0$ and $z=2d$ yields a matrix equation of the form $M\mathbf{v}=0$ where $M$ is an 8-by-8 matrix and $\mathbf{v}=(A_1^\mathrm{TM},A_{2-}^\mathrm{TM},A_{2+}^\mathrm{TM},A_3^\mathrm{TE},A_1^\mathrm{TE},A_{2-}^\mathrm{TE},A_{2+}^\mathrm{TE},A_3^\mathrm{TE})$. The elements of the matrix $M$ are 
\begin{equation*}
	M = 
	\left(
	\begin{array}{c|c}
		M_{11} & M_{12} \\\hline  M_{21} & M_{22}
	\end{array}
	\right)
\end{equation*}

\begin{equation*}
	M_{11}=
	\begin{pmatrix}
		\frac{\sigma_\chi^1}{2} & \frac{\sigma_\chi^1}{2} & \frac{\sigma_\chi^1}{2} & 0 \\
		-1 & 1 & 1 & 0\\
		-\frac{\sigma_e^1}{2}+\tilde{\epsilon}_1 & -\frac{\sigma_e^1}{2}+\tilde{\epsilon}_2 & -\frac{\sigma_e^1}{2}-\tilde{\epsilon}_2 & 0 \\
		\tilde{\epsilon}_1\frac{\sigma_\chi^1}{2} & -\tilde{\epsilon}_2\frac{\sigma_\chi^1}{2} & \tilde{\epsilon}_2\frac{\sigma_\chi^1}{2} & 0
	\end{pmatrix}
\end{equation*}
\begin{equation*} 
	M_{12} = 
	\begin{pmatrix}
		-i & i & i & 0 \\
		-i\frac{\sigma_\chi^1}{2} & -i\frac{\sigma_\chi^1}{2} & -i\frac{\sigma_\chi^1}{2} & 0 \\
		i\frac{\sigma_\chi^1}{2\tilde{\mu}_1} &  -i\frac{\sigma_\chi^1}{2\tilde{\mu}_2} & i\frac{\sigma_\chi^1}{2\tilde{\mu}_2} & 0 \\
		i\left(\frac{\sigma_e^1}{2}-\frac{1}{\tilde{\mu}_1}\right) & i\left(\frac{\sigma_e^1}{2}-\frac{1}{\tilde{\mu}_2}\right) & i\left(\frac{\sigma_e^1}{2}+\frac{1}{\tilde{\mu}_2}\right) & 0
	\end{pmatrix}
\end{equation*}

\begin{equation*}
	M_{21}=
	\begin{pmatrix}
		0 & \frac{\sigma_\chi^2}{2}e^{-2q'_2d} & \frac{\sigma_\chi^2}{2}e^{2q'_2d} & \frac{\sigma_\chi^2}{2}e^{-2q'_3d} \\
		0 & -e^{-2q'_2d} & -e^{2q'_2d} & e^{-2q'_3d} \\ 
		0 & \left(-\frac{\sigma_e^2}{2}-\tilde{\epsilon}_2\right)e^{-2q'_2 d} & \left(-\frac{\sigma_e^2}{2}+\tilde{\epsilon}_2\right)e^{2q'_2 d} & \left(-\frac{\sigma_e^2}{2}+\tilde{\epsilon}_3\right)e^{-2q'_3 d} \\ 
		0 & -\tilde{\epsilon_2}\frac{\sigma_\chi^2}{2}e^{-2q'_2d} & \tilde{\epsilon_2}\frac{\sigma_\chi^2}{2}e^{2q'_2d} & -\tilde{\epsilon_3}\frac{\sigma_\chi^2}{2}e^{-2q'_3d}
	\end{pmatrix}
\end{equation*}

\begin{equation*}
	M_{22} = 
	\begin{pmatrix}
		0 & -i e^{-2q'_2d} & -i e^{2q'_2d} & i e^{-2q'_3d} \\ 
		0 & -i\frac{\sigma_\chi^2}{2} e^{-2q'_2d} & -i\frac{\sigma_\chi^2}{2} e^{2q'_2d} & -i\frac{\sigma_\chi^2}{2} e^{-2q'_3d} \\ 
		0 & -i\frac{\sigma_\chi^2}{2\tilde{\mu}_2}e^{-2q'_2d} & i\frac{\sigma_\chi^2}{2\tilde{\mu}_2}e^{2q'_2d} & -i\frac{\sigma_\chi^2}{2\tilde{\mu}_3}e^{-2q'_3d} \\ 
		0 & i\left(\frac{\sigma_e^2}{2}+\frac{1}{\tilde{\mu}_2}\right)e^{-2q'_2d} & i\left(\frac{\sigma_e^2}{2}-\frac{1}{\tilde{\mu}_2}\right)e^{2q'_2d} & 
		i\left(\frac{\sigma_e^2}{2}-\frac{1}{\tilde{\mu}_3}\right)e^{-2q'_3d}
	\end{pmatrix}
\end{equation*}
where we have defined $\tilde{\epsilon}_j=i\omega\epsilon_j\epsilon_0/q'_j$ and $\tilde{\mu}_j=i\omega\mu_j\mu_0/q'_j$. The dispersion for the plasmon can be found from $\det(M)=0$. The plasmon field can be constructed from the vector $\mb{v}$.

\begin{figure}
    \centering
    \includegraphics{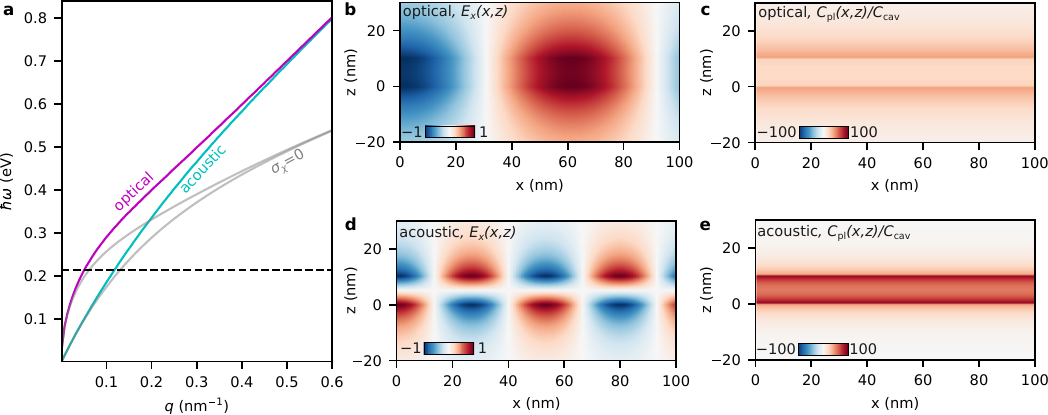}
    \caption{Dispersion, electric field, and optical chirality of the plasmon for a $\sigma_\chi^1=\sigma_\chi^2$ bilayer. The layer separation is set to $2d=10$nm and the chiral conductivity of both layers is $\sigma_\chi=0.01$. \textbf{a}, Dispersion of both the optical and acoustic modes are shown. The grey lines show the optical and acoustic modes for the bilayer with $\sigma_\chi=0$. The dashed line indicates the frequency of interest $\hbar\omega_0=0.215$meV. \textbf{b-e},The $x$-component of the electric field $E_x$ and the optical chirality enhancement $C_{\mathrm{pl}}/C_{\mathrm{cav}}$ for the optical and acoustic modes. All fields are plotted for the frequency of interest $\omega_0$.}
    \label{fig:bilayer_chiral}
\end{figure}

\begin{figure}
    \centering
    \includegraphics{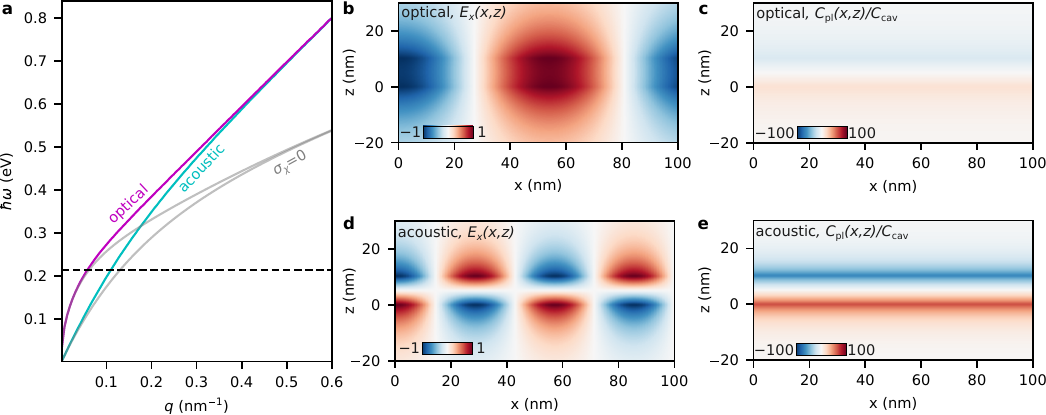}
    \caption{Dispersion, electric field, and optical chirality of the plasmon for a $\sigma_\chi^1=-\sigma_\chi^2$ bilayer. The layer separation is set to $2d=10$nm and the chiral conductivity of each layer is $\sigma^1_\chi=0.01=-\sigma_\chi^2$. \textbf{a}, Dispersion of both the optical and acoustic modes are shown. The grey lines show the optical and acoustic modes for the bilayer with $\sigma_\chi=0$. The dashed line indicates the frequency of interest $\hbar\omega_0=0.215$meV. \textbf{b-e},The $x$-component of the electric field $E_x$ and the optical chirality enhancement $C_{\mathrm{pl}}/C_{\mathrm{cav}}$ for the optical and acoustic modes. All fields are plotted for the frequency of interest $\omega_0$.}
    \label{fig:bilayer_achiral}
\end{figure}

The plasmon dispersion, electric field, and optical chirality for the bilayer case with $\sigma_\chi^1=\sigma_\chi^2$ is shown in \cref{fig:bilayer_chiral}. The handedness-preserving reflector would retrieve the acoustic mode of this setup. Alternatively, \cref{fig:bilayer_achiral} shows the case of $\sigma_\chi^1=-\sigma_\chi^2$, which corresponds to a reflector that does not preserve handedness. In both cases we assume $\sigma_e^1=\sigma_e^2=iD/\omega$ where the Drude weight is set to $D=\sigma_0\omega_0$. 

\FloatBarrier
\bibliography{references}